\documentclass[twocolumn,showpacs,preprintnumbers,amsmath,amssymb]{revtex4}

\usepackage{graphicx}
\usepackage{dcolumn}
\usepackage{bm}

\begin{document}


\title{Can One Extract the Neutrino Oscillation Signature from the SuperKamiokande Experiment ?  \\ An Analysis of Neutrino Events Occurring outside the Detector}

\author{N. Takahashi}
 \email{taka@cc.hirosaki-u.ac.jp}
 \affiliation{Faculty of Science and Technology, Hirosaki University, 
  036-8561, Hirosaki, Japan}
\author{A. Misaki}
 \affiliation{Advanced Research Institute for Science and Engineering, 
  Waseda University, 169-0092, Tokyo, Japan}
 \affiliation{Innovative Research Organization for the New Century, 
  Saitama University, 338-8570, Saitama, Japan}

\date{\today}

\begin{abstract}
The SuperKamiokande group assert that they have found an oscillatory signature
in atmospheric neutrinos through the analysis of 
\textit{Fully Contained Events} and \textit{Partially Contained Events}. 
We have performed an $L/E$ (length/energy) analysis of 
\textit{Upward Through-Going Muon Events} and 
\textit{Stopping Muon Events} in a numerical
computer simulations both with and without neutrino oscillations but 
were unable to find an oscillatory signature. We give likely explanations
for the absence of the oscillatory signature in our simulations
and its apparent presence in the SuperKamiokande data.
\end{abstract}

\maketitle

The SuperKamiokande group (hereafter SK) have reported 
the presence of an oscillatory signature
in atmospheric neutrino 
data through an $L/E_{\nu}$ analysis
of neutrino events occurring inside the detector, using 
\textit{Fully Contained Events} and \textit{Partially Contained Events} 
with energies ranging from several hundred MeV to several GeV
[1].  Recently, it has become  recognized that the
scattering angle of the emitted lepton greatly influences the
estimation of the direction of the incident neutrino, which is directly
connected with the determination of $L$ in the energy region of
several hundred MeV to several GeV
[2].  In the 
SK analysis, it is necessary for $L$ to be decided more accurately,
because the oscillatory signature is strongly sensitive to
$L/E_{\nu}$.

For the $L/E_{\nu}$ analysis in the SK experiment, it is more 
appropriate
to analyze neutrino events occurring outside the detector, such as
\textit{Upward Through-Going Muon Events} and 
\textit{Stopping Muon Events} whose energies range from several 
GeV to several
hundreds of GeV, because the scattering angle of the emitted lepton in
the neutrino reactions can be neglected at such higher energies and
consequently the direction of the emitted muon can be approximated as
that of the neutrino events, which results in higher accuracy in the
determination of $L$.
In this paper, we report the result of an analysis for an oscillatory
signature in \textit{Upward Through-Going Muon Events} and
\textit{Upward Stopping Muon Events} in a virtual SuperKamiokande
experiment by numerical computer simulations. The particles which
produce such events are regarded exclusively as muons (or muon neutrinos) 
due to
their long flight length compared to the cascade shower initiated by
electrons (or electron neutrinos).

In both the present analysis of the neutrino events and that of SK,
a careful examination on the validity of the technique for the Monte
Carlo Method utilized is vital, and we therefore 
start by explaining the procedures of the Monte Carlo Method we utilize in
the present report. Our simulation can be regarded as a
\textit{Time Sequential Simulation}, while the SK groups is a 
\textit{Detector Simulation}.

We start our simulation,
which is schematically shown in Figure~1, 
with the atmospheric neutrino energy
spectrum at the opposite side of the Earth to the detector. 
We utilize Honda's spectrum [3] as the incident atmospheric neutrino
energy spectrum. 
We adopt a maximum energy of 1000\,GeV, and therefore 
the maximum energy of the
muon emitted from the neutrino interaction, here, is 1000\,GeV.  
We calculate the range fluctuation of 
1000\,GeV muons by
the exact Monte Carlo method, taking into the physical
processes concerned --- bremsstrahlung, direct pair
production, nuclear interaction and ionization loss --- and show
the result in Figure~2.

We conclude from Figure~2 that it is sufficient to consider the
neutrino events for \textit{Upward Through-Going Muon Events} and
\textit{Stopping Muon Events}, which are generated in the region 
within 400000\,g/cm$^{2}$ of the SK detector 
(less than 2000 meters), because neutrino interactions 
further than 2000\,m from
the detector could not contribute physical events into the
detector.

Then, we define $N_{int}( E_{\nu},t,cos\theta_{\nu})dt$, the
interaction neutrino energy spectrum at depth $t$ from the
detector underground for the incident neutrino with energy $E_{\nu}$
from the zenith angle $\theta_{\nu}$, in the following.

\begin{figure}
\includegraphics[width=0.8\linewidth]{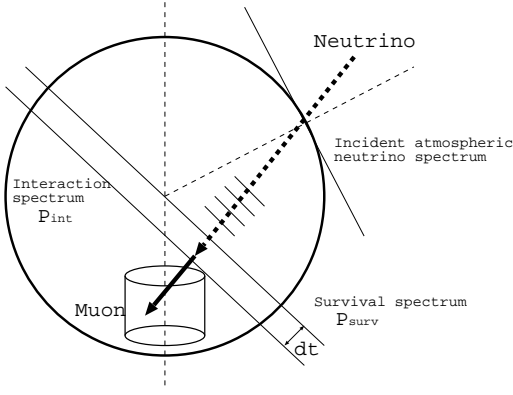}
\caption{\label{fig:1} Schematic illustration of the experiment.}
\end{figure}

\begin{figure}
\includegraphics[width=0.8\linewidth]{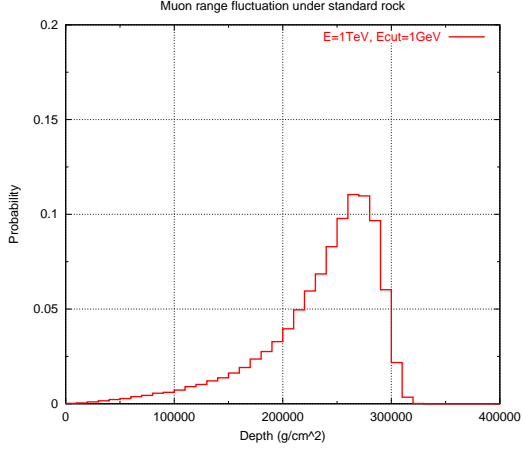}
\caption{\label{fig:2} Range Energy Fluctuation of 1000\,GeV muons.}
\end{figure}

\begin{figure}
\includegraphics[width=0.75\linewidth]{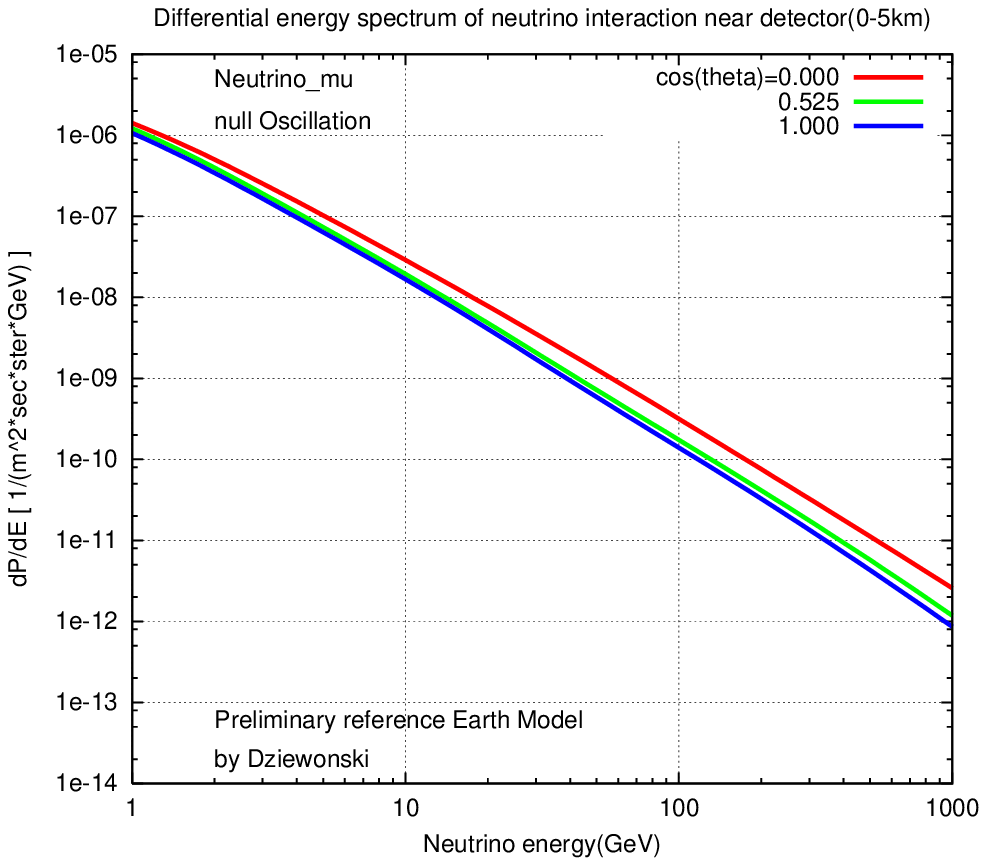}
\caption{\label{fig:3}  Interaction energy spectrum without 
neutrino oscillation.}
\end{figure}

\begin{figure}
\includegraphics[width=0.75\linewidth]{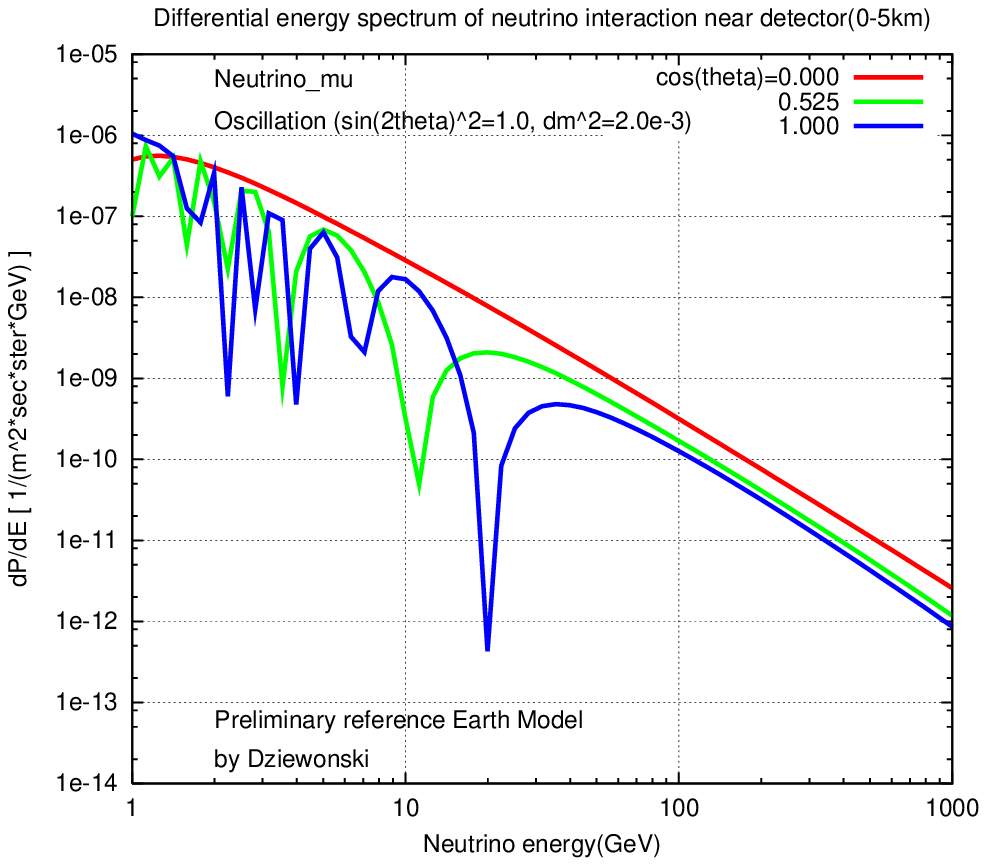}
\caption{\label{fig:4} Interaction energy spectrum with 
neutrino oscillation.}
\end{figure}

\begin{eqnarray}
\lefteqn{N_{int}(E_{\nu},t,cos\theta_{\nu})dt
=N_{sp}(E_{\nu},cos\theta_{\nu}) \times } \hspace{8cm} \nonumber\\
\left(1-\frac{dt}{\Lambda_{1}( E_{\nu},t_{1},\rho_{1})}\right)
\times \left(1-\frac{dt}{\Lambda_{2}( E_{\nu},t_{2},\rho_{2})}\right)
\times \cdot \cdot \nonumber\\
\cdot \cdot 
\times \left(1-\frac{dt}{\Lambda_{n-1}( E_{\nu},t_{n-1},\rho_{n-1})}\right)
\times \left(\frac{dt}{\Lambda_{n}( E_{\nu},t_{n},\rho_{n})}\right)\nonumber\\
\end{eqnarray}

Here, $\Lambda_{i}(E_{\nu},t_{i},\rho_{i})$ is given by

\[
\frac{1}{\Lambda_{i}(E_{\nu},t_{i},\rho_{i})}
=\frac{1}{\lambda_{i}(E_{\nu},t_{i},\rho_{i})}
+\frac{1}{\lambda_{i}(E_{\bar{\nu}},t_{i},\rho_{i})}
\]

where $\lambda_{i}(E_{\nu},t_{i},\rho_{i})$ denotes the mean free
path of the neutrino with energy $E_{\nu}$ at the distance $t_{i}$
from the opposite surface of the Earth whose density is $\rho_{i}$ and
$\lambda_{i}(E_{\bar{\nu}},t_{i},\rho_{i})$ denotes the corresponding
mean free path of the anti-neutrino whose energy is given by
$E_{\bar{\nu}}$. These mean free paths are calculated from the deep
inelastic scattering cross sections 
[4].  The density of the Earth is adopted from the Preliminary Earth
Model for the density profile
[5].

The distribution functions for $L/E_{\nu}$,
obtained using Eq.(1), are given in Eq.(2) and Eq.(3)
in the cases without neutrino oscillation and with
neutrino oscillation, respectively.

\begin{eqnarray}
\lefteqn{N^{null-osc}_{\nu}\left(\frac{L}{E_{\nu}},cos\theta_{\nu}\right)d\left(\frac{L}{E_{\nu}}\right)} \hspace{7cm} \nonumber\\
= N_{\nu}(E_{\nu},L,cos\theta_{\nu})d\left(\frac{L}{E_{\nu}}\right)
\end{eqnarray}

\begin{eqnarray}
\lefteqn{N^{osc}_{\nu}\left(\frac{L}{E_{\nu}},cos\theta_{\nu}\right)d\left(\frac{L}{E_{\nu}}\right)} \hspace{7cm} \nonumber\\
= N_{\nu}(E_{\nu},L,cos\theta_{\nu})P(\nu_{\mu}\longrightarrow \nu_{\mu})d\left(\frac{L}{E_{\nu}}\right)
\end{eqnarray}

where $P(\nu_{\mu}\longrightarrow \nu_{\mu})$ denotes the survival
probability for a neutrino in the presence of neutrino oscillations,
which is given as

\begin{eqnarray}
\lefteqn{P(\nu_{\mu}\longrightarrow \nu_{\mu})} \hspace{7cm} \nonumber\\
=1-sin^{2}2\theta sin^{2}\left( \frac{1.27\Delta m^{2}(eV^{2})L(km)}{E_{\nu }(GeV)} \right)
\end{eqnarray}

Here, we adopt $sin^{2}2\theta=1.00$ and 
$\Delta m^{2}=2\times 10^{-3} eV^{2}$, as obtained from the SK experiment.

In Figures~3 and 4, we give the interaction neutrino energy
spectrum without, and with, neutrino oscillations
as defined by Eq.(2) and Eq.(3), respectively.  It is clear
from Figure~4 that the effect of the neutrino oscillation for the SK
parameters does not appear in the horizontal direction,
$cos\theta_{\nu}=0.0$, due to short path length for the traversed
neutrino while the effect clearly appears the vertical case,
$cos\theta_{\nu}=1.0$

For \textit{Upward Through-Going Muon Events} and 
\textit{Stopping Muon Events}, 
we can assume that the direction of the incident
neutrino is the same as that of the emitted muon, because the
scattering angle of the emitted muon can be neglected due to its
high energy. The simulation procedures for the events concerned for
a given zenith angle of the incident neutrinos are as follows.

\begin{figure}
\includegraphics[width=0.75\linewidth]{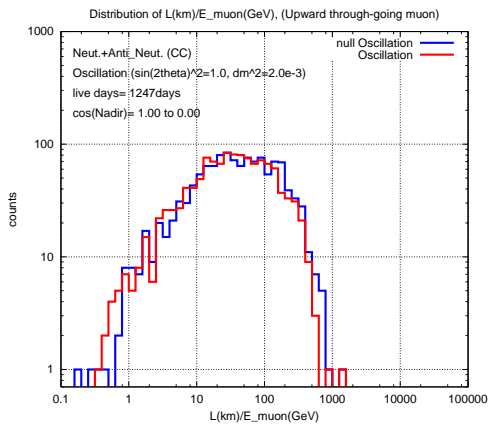}
\caption{\label{fig:5}  Distribution of $L/E_{\mu}$ 
 for \textit{Upward Through-Going Muon Events}.}
\end{figure}

\begin{figure}
\includegraphics[width=0.75\linewidth]{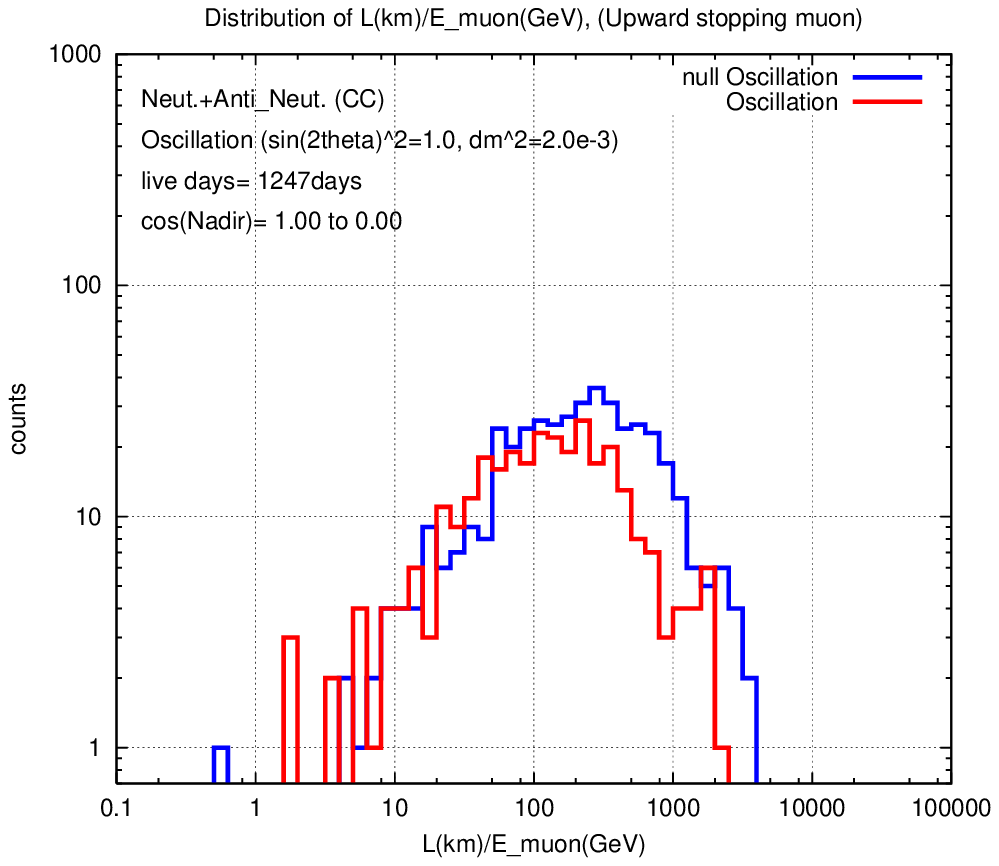}
\caption{\label{fig:6} Distribution of $L/E_{\mu}$ 
for \textit{Upward Stopping Muon Events}.}
\end{figure}

\underline{Procedure A:} 
Using $\xi$, a uniform random number
between 0 and 1, we sample $E_{\nu(\bar{\nu})}$, the energy of the
incident (anti-)neutrino, which is obtained from the following equation:

\begin{eqnarray}
\xi=\frac{\int^{E_{\nu}}    _{E_{\nu,min}}N_{int}(E_{\nu},t,cos\theta_{\nu})dt}
         {\int^{E_{\nu,max}}_{E_{\nu,min}}N_{int}(E_{\nu},t,cos\theta_{\nu})dt}
\end{eqnarray}

\underline{Procedure B:} 
For $E_{\nu(\bar{\nu})}$, the energy of
the (anti-)neutrino obtained by Procedure~A, we define

\[
\xi_{1}=\frac{\sigma(E_{\nu})}{\sigma(E_{\nu})+\sigma(E_{\bar{\nu}})}
\]

where $\sigma(E_{\nu})$ and $\sigma(E_{\bar{\nu}})$ denote the total
cross-section of the neutrino and the anti-neutrino, respectively.
Then we sample the random number $\xi$ again and if
$\xi \le \xi_{1}$ we take the incident lepton to be a neutrino,
otherwise we take it to be an anti-neutrino.

\underline{Procedure C:} 
We decide the interaction point of the
(anti-)neutrino event determined by Procedure~B in the range (0, 2000) meters.
This is the distance from the detector to the interaction
point and is obtained simply
by sampling a uniform random number between (0,1)
as this range is many orders of magnitude smaller than 
the mean free path of the neutrino concerned.

\underline{Procedure D:} 
We sample $E_{\mu}$, the energy of the
(anti-)muon emitted for the deep inelastic scattering by using the
uniform random number between (0,1), which is logically same as in
Eq.(5).

\underline{Procedure E:} 
For the (anti-)muon whose energy and production
point is determined from Procedures~C and D, we examine the behavior
of the trajectory of the (anti-)muon toward the SK detector in a 
stochastic manner.
Namely, each individual muon is pursued by taking into consideration
bremsstrahlung, direct pair production, nuclear interaction, 
and ionization loss without
utilizing the average behavior of the muon concerned.  As the result,
we determine which category each individual muons falls into:
[a] stopping before it reaches the detector, [b] stopping inside
detector, or [c] passing through the detector.

We repeat Procedures~A to E and obtain the neutrino
events concerned for a given live time for the real experiment.  In our
computer numerical simulation, we have accumulated the events
concerned which correspond to the real live time for SK. For each
neutrino event, we know $E_{\nu}$, the energy of the parent
neutrino, $E_{\mu}$, the energy of the daughter (anti-)muon,
$cos\theta_{\nu}$ ($cos\theta_{\mu}$), the direction of both the
incident (anti-)neutrino and the emitted (anti-)muon, and L, the distance
between the interaction point of the neutrino events and
the opposite side of the Earth.  

As only $E_{\mu}$ can be measured in the actual SK experiment, and
not $E_{\nu}$, we
give the frequency of the number of the events as a function of
$L/E_{\mu}$ for \textit{Upward Through-Going Muon} in Figure~5 and
the corresponding quantity for \textit{Stopping Muon Events} in Figure~6 
in both the cases without and with neutrino
oscillations. It is clear from those figures that no oscillatory
signature is apparent, with almost no difference between the cases
with and without oscillation when plotted against the measured SK parameter.
Of course, according to the logic adopted
by SK, the oscillatory signature should appear in the function of
$L/E_{\nu}$, not $L/E_{\mu}$.

\begin{figure}
\includegraphics[width=0.75\linewidth]{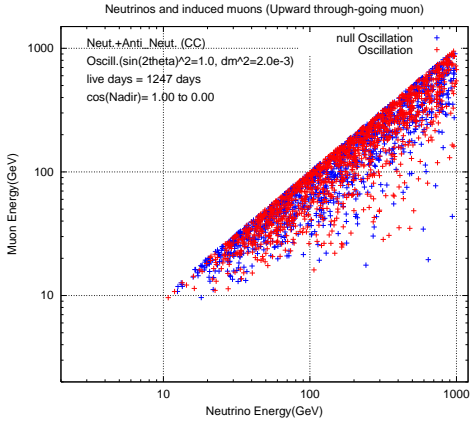}
\caption{\label{fig:7} $E_{\nu}$ vs $E_{\mu}$ scatter plot for 
\textit{Upward Through-Going Muon Events}.}
\end{figure}

\begin{figure}
\includegraphics[width=0.75\linewidth]{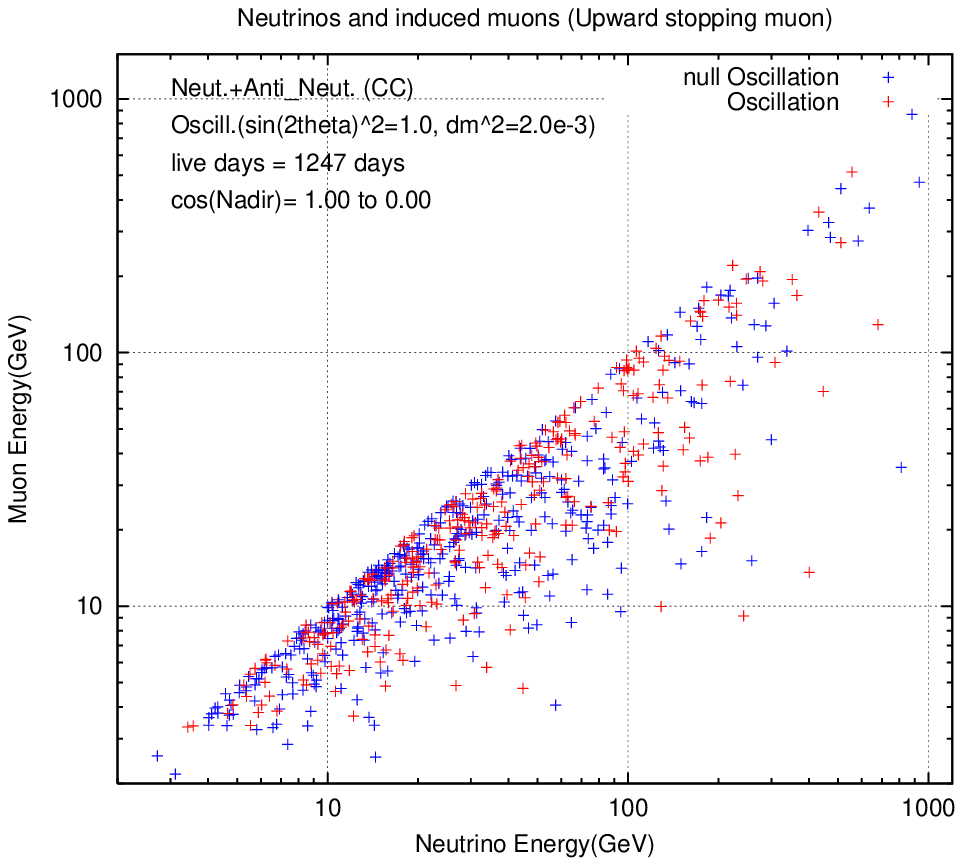}
\caption{\label{fig:8} $E_{\nu}$ vs $E_{\mu}$ scatter plot for 
\textit{Upward Stopping Muon Events}.}
\end{figure}

\begin{figure}
\includegraphics[width=0.75\linewidth]{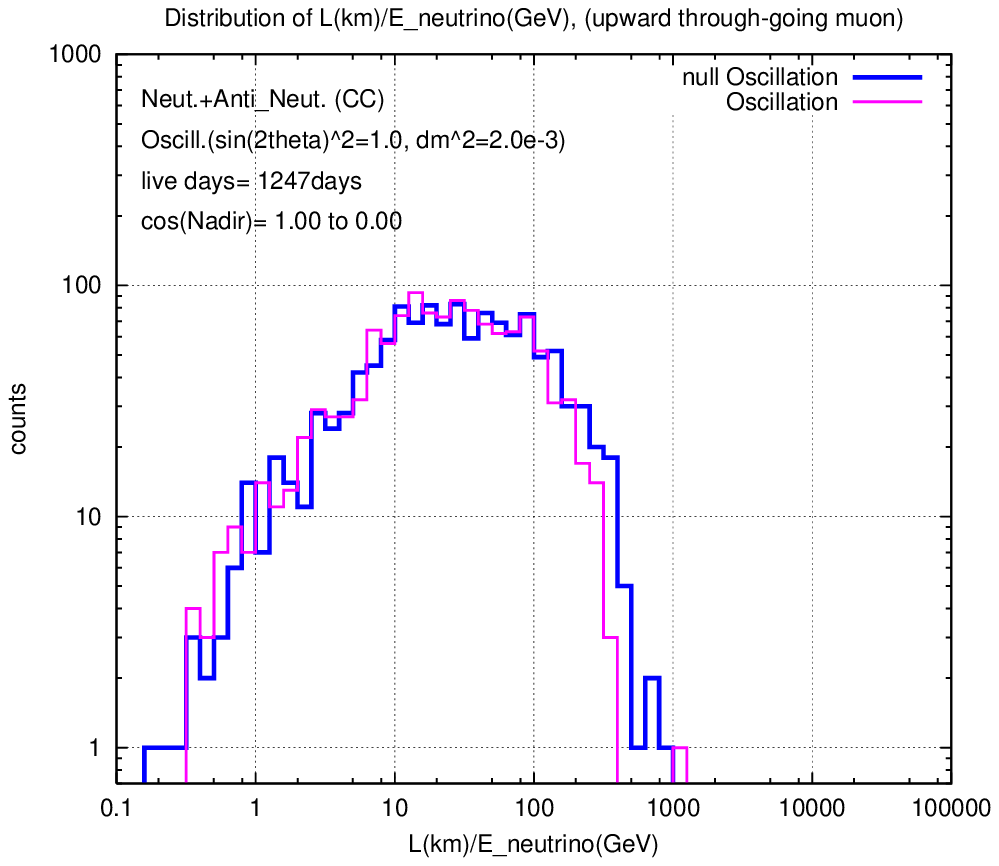}
\caption{\label{fig:9} Distribution of $L/E_{\nu}$ 
for \textit{Upward Through-Going Muon Events}.}
\end{figure}

\begin{figure}
\includegraphics[width=0.75\linewidth]{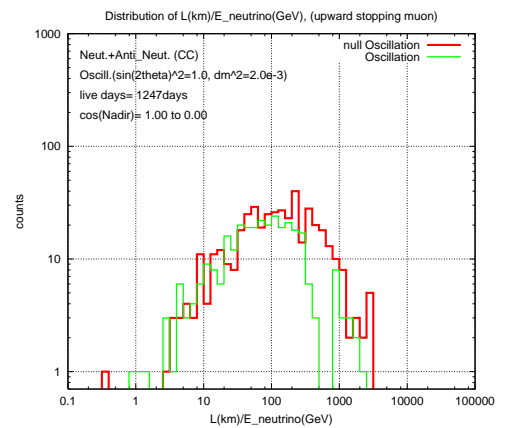}
\caption{\label{fig:10} Distribution of $L/E_{\nu}$ 
for \textit{Upward Stopping Muon Events}.}
\end{figure}

The energy of the emitted lepton for a given energy of the neutrino is
generated in the deep elastic scattering, producing a wider energy
distribution.  In Figures~7 and 8, we give the scatter plots
between $E_{\nu}$ and $E_{\mu}$  for 
\textit{Upward Through-Going Muon Events} and
\textit{Stopping Muon Events}, respectively. It is clear
from the figures that the fluctuations in the energy
distributions are not small. The density of points in the figures is
proportional to the numbers of events.

In Figures~9 and 10, we give the corresponding distributions for
$L/E_{\nu}$ to Figures~5 and 6 for $L/E_{\mu}$. From the comparison of
Figures~5 and 6 with Figures~9 and 10, it can be concluded that
oscillatory signatures are not observed in both $L/E_{\nu}$ and
$L/E_{\mu}$ distributions. Also, there are only small differences 
between the $L/E$
distributions with and without neutrino distribution, as seen
in Figures~5 and 6 and in Figures~9 and 10, which is 
consistent with our previous
conclusion that there is no evidence for neutrino oscillation 
through the analysis
of the zenith angle distribution for 
\textit{Upward Through-Going Muon Events} and 
\textit{Stopping Muon Events}
[6].

In Figures~11 and 12, we give the ratios of ``with oscillation'' 
to ``without oscillation''  as the function of $L/E_{\nu}$ for 
\textit{Upward Through-Going Muon Events} and 
\textit{Stopping Muon Events}, respectively.  
Generally, the fact that fluctuations are rather larger comes from 
relatively small number of events, and further,
the fluctuations are larger in \textit{Upward Stopping Muon Events}
than in \textit{Upward Through-Going Muon Events}, as they must be. In
Figures~13 and 14, we give the corresponding quantities for
$L/E_{\mu}$. Comparing Figures~11 and 12 with Figures~13 and 14,
the fluctuations are larger in the latter than in the former, which is 
easily understandable when considering the energy distribution of 
$E_{\mu}$ for a definite $E_{\nu}$. 
It is clear from these figures that it is exceedingly difficult to 
extract some positive evidence for neutrino oscillations. 
In Figures~15 and 16, we give scatter plots between $L$ and $E_{\nu}$ with and 
without neutrino oscillation for \textit{Upward Through-Going Muon 
Events} and \textit{Stopping Muon Events}, respectively. Also, we 
give the corresponding plots between $L$ and $E_{\mu}$ in Figures~17 
and 18.

\begin{figure}
\includegraphics[width=0.75\linewidth]{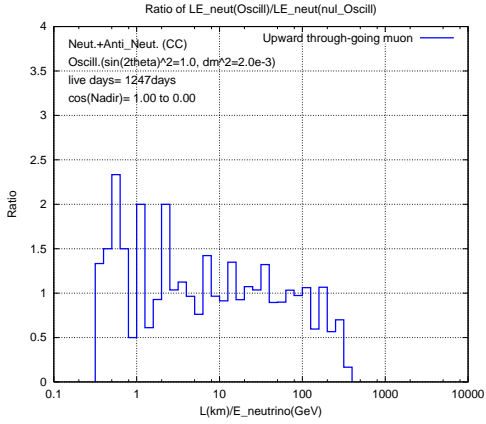}
\caption{\label{fig:11} The ratio of ``oscillation'' to ``null oscillation'' of  
$L/E_{\nu}$ distribution for \textit{Upward Through-Going Muon Events}.}
\end{figure}

\begin{figure}
\includegraphics[width=0.75\linewidth]{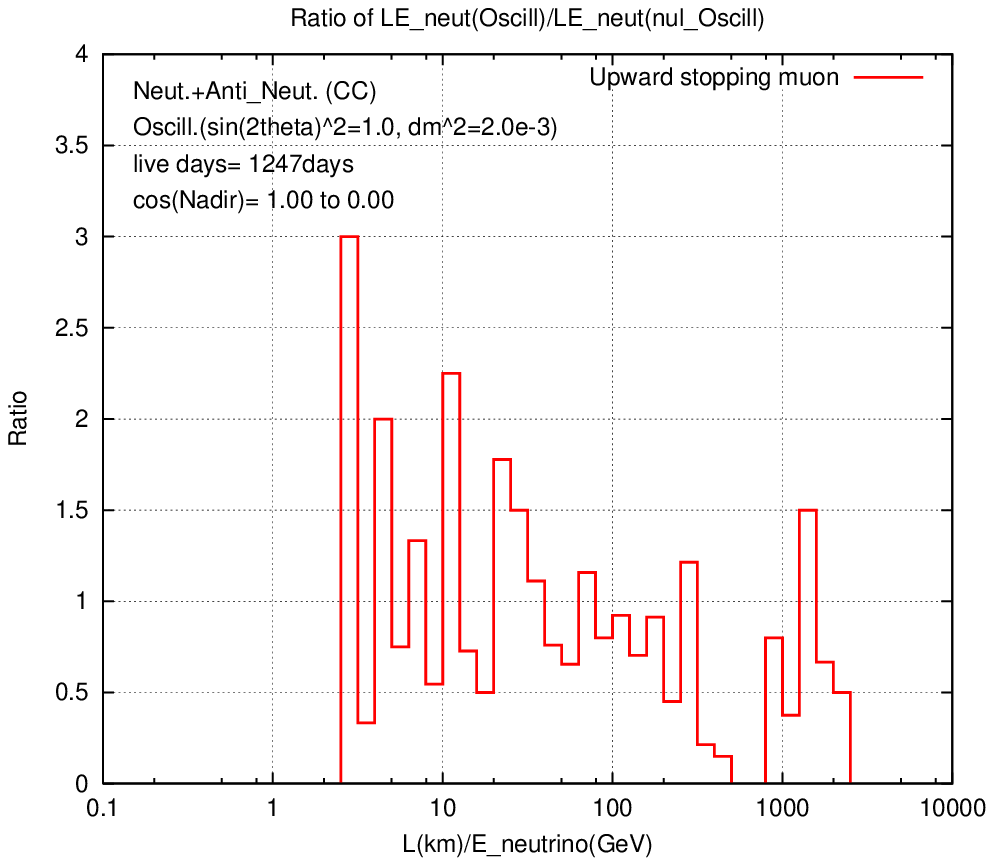}
\caption{\label{fig:12} The ratio of ``oscillation'' to ``null oscillation'' of 
$L/E_{\nu}$ distribution for \textit{Upward Stopping Muon Events}.}
\end{figure}

\begin{figure}
\includegraphics[width=0.75\linewidth]{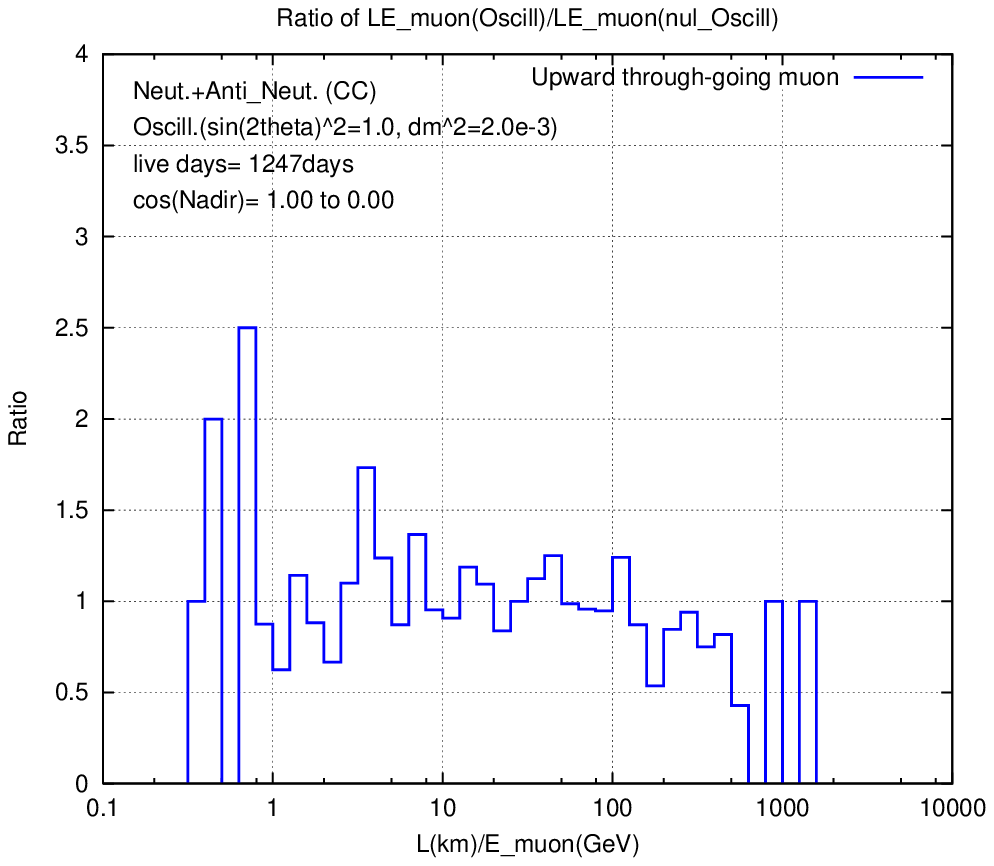}
\caption{\label{fig:13} The ratio of ``oscillation'' to ``null oscillation'' of 
$L/E_{\mu}$ distribution for \textit{Upward Through-Going Muon Events}.}
\end{figure}

\begin{figure}
\includegraphics[width=0.75\linewidth]{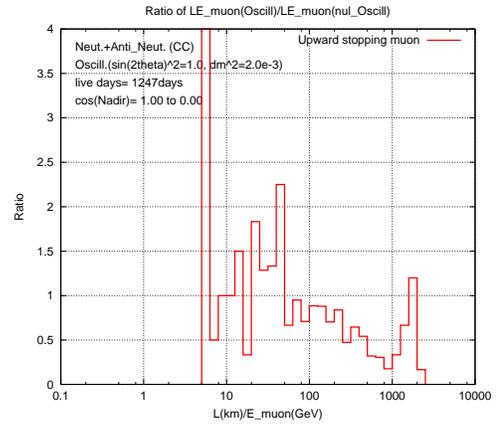}
\caption{\label{fig:14} The ratio of ``oscillation'' to ``null oscillation'' of 
$L/E_{\mu}$ distribution for \textit{Upward Stopping Muon Events}.}
\end{figure}

Here, we comment on the evidence for an oscillatory signature claimed by SK. 
In our opinion, it is practically impossible to observe such an oscillatory 
signature in atmospheric neutrino oscillation for the following reasons:

(i) The $L/E$ distribution for the muon events occurring inside the
detector in the SK experiment, $N(L/E_{\nu})d(L/E_{\nu})$ is given as,

\begin{eqnarray}
\lefteqn{N\left(\frac{L}{E_{\nu}}\right)d\left(\frac{L}{E_{\nu}}\right) =} \hspace{9cm} \nonumber\\
d\left(\frac{L}{E_{\nu}}\right)\int dcos\theta_{\nu} D_{nsp}(E_{\nu},cos\theta_{\nu},L(cos\theta_{\nu})) 
\hspace{1.5cm} \nonumber\\
\times P(\nu_{\mu}\longrightarrow \nu_{\mu}) \int^{E_{\mu,max}}_{E_{\mu,min}} \sigma_{\nu \rightarrow \mu}(E_{\nu},E_{\mu}) dE_{\mu}
\hspace{2.0cm}
\end{eqnarray}

where $D_{nsp}(E_{\nu},cos\theta_{\nu},L(cos\theta_{\nu}))$ is the
differential neutrino incident energy spectrum at the given zenith
angle, $\theta_{\nu}$, and $L$, which is also the function of
$cos\theta_{\nu}$.  $P(\nu_{\mu}\longrightarrow \nu_{\mu}) $ denotes
the probability function for the neutrino oscillation and 
$\sigma_{\nu \rightarrow \mu}(E_{\nu},E_{\mu}) dE_{\mu} $ 
denotes the differential
cross-section for the incident neutrino to produce the emitted muon
inside the detector.  The character of a oscillatory signature
obtained by SK comes exclusively from the part of
$P(\nu_{\mu}\longrightarrow \nu_{\mu})$ in Eq.(6). However,
$P(\nu_{\mu}\longrightarrow \nu_{\mu})$ in Eq.(6) is imbedded into the
steep neutrino energy spectra 
like that shown in Figure~4 and
is itself a continuously and strongly varying function with L.
As a result, it can not be separated from the neutrino
energy spectrum.

\begin{figure}
\includegraphics[width=0.8\linewidth]{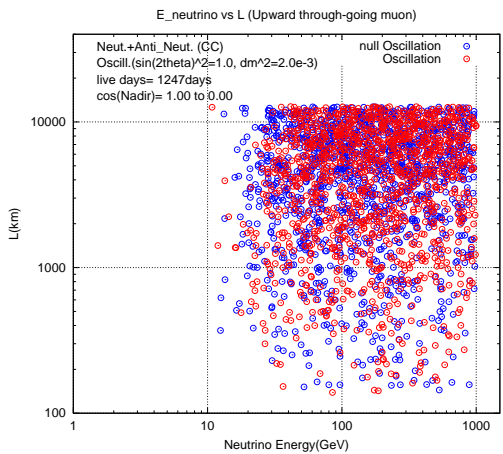}
\caption{\label{fig:15} 
$L$ vs $E_{\nu}$ scatter plot for \textit{Upward Through-Going Muon Events}.}
\end{figure}

\begin{figure}
\includegraphics[width=0.8\linewidth]{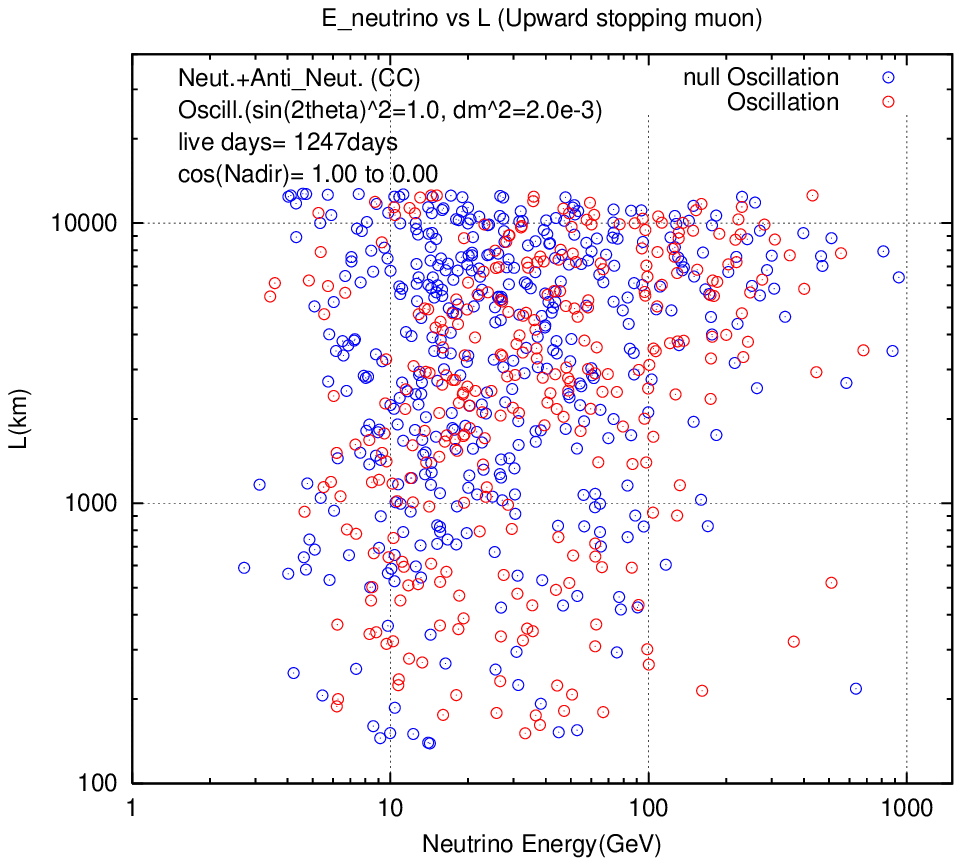}
\caption{\label{fig:16} 
$L$ vs $E_{\nu}$ scatter plot for \textit{Upward Stopping Muon Events}.}
\end{figure}

\begin{figure}
\includegraphics[width=0.8\linewidth]{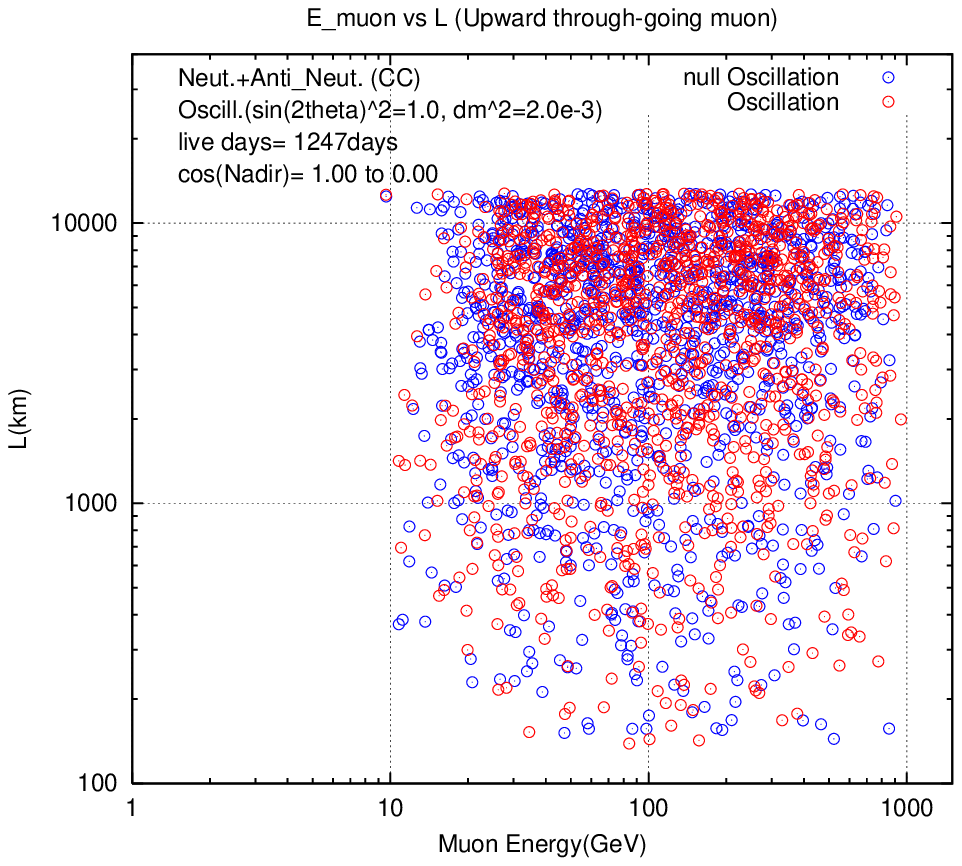}
\caption{\label{fig:17} 
$L$ vs $E_{\mu}$ scatter plot for \textit{Upward Through-Going Muon Events}.}
\end{figure}

\begin{figure}
\includegraphics[width=0.8\linewidth]{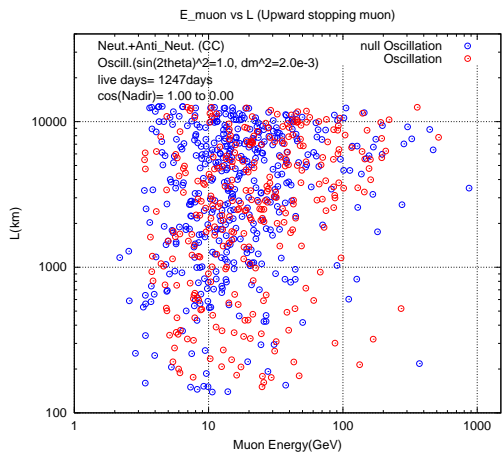}
\caption{\label{fig:18} 
$L$ vs $E_{\mu}$ scatter plot for \textit{Upward Stopping Muon Events}.}
\end{figure}

(ii) The SK Detector Simulation assumes that the direction of the 
incident neutrino is the same as that of the emitted lepton 
[7,8]. Thus, SK could treat neither the effect of the azimuthal angle of
the emitted lepton nor the effect of the backscattering over the
zenith angle of the emitted lepton and consequently could not determine $L$
reliably, which is directly connected with $N(L/E_{\nu})d(L/E_{\nu})$.
More concretely, under the combination of the two fundamental
parameters in the neutrino oscillation derived by SK, about $40\%$ of
the upward going muons originate from downward going neutrinos, while
about $10\%$ of the downward going muons originate from upward going
neutrinos [9]. 
Such mutual mixing in the direction of the incident
neutrinos may be maximized near $ L/E_{\nu}= 150 \sim 500$ (km/GeV), 
where one find the dip-like phenomena of the $N(L/E_{\nu})d(L/E_{\nu})$ 
in Figure~3 in the SK paper 
[1]. If we determine the direction of
the incident neutrino correctly, then the dip will disappear and
$N(L/E_{\nu})d(L/E_{\nu})$ will show the something like behavior given in
Figures~9 and 10.  Our subsequent paper on the $L/E$ analysis for
\textit{Fully Contained Events} and 
\textit{Partially Contained Events} will be published elsewhere.



{\large References} \\
$[1]$ Y.\ Ashie {\it et al.}, Phys.\ Rev.\ Lett.\ {\bf 93}, 101801 (2004).\\
$[2]$ E.\ Konishi {\it et al.}, astro-ph/0406497. \\
$[3]$ M.\ Honda {\it et al.}, Phys.\ Rev. D {\bf 52}, 4985 (1996). \\
$[4]$ R.H.\ Gandhi {\it et al.}, Astropart.\ Phys.\ {\bf 5}, 81 (1996). \\
$[5]$ A.\ Dziewonski, 
in \textit{The Encyclopedia of Solid Earth Geophysics} 
ed. D.E. James (Van Nostrand Reinhold, New York) (1989). \\
$[6]$ N.\ Takahashi {\it et al.}, Proc.\ of the 28-th ICRC, 1275 (2003).\\
$[7]$ T.\ Kajita and Y.\ Totsuka, Rev.\ Mod.\ Phys.\ {\bf 73}, 85 (2001). \\
$[8]$ M.\ Ishitsuka, Ph.D thesis, University of Tokyo, (2004). \\
$[9]$ E.\ Konishi {\it et al.}, to be submitted to hep-ex. \\

\end{document}